\documentclass[10pt,showpacs,amsmath,amssymb,floatfix,superscriptaddress,longbibliography,twocolumn,eqsecnum]{revtex4-2}
\usepackage{amsmath}
\usepackage{physics}
\usepackage{graphicx}
\usepackage{wrapfig}
\usepackage[usenames,dvipsnames,svgnames,table]{xcolor}
\usepackage{tcolorbox}
\usepackage{float}
\usepackage{booktabs}
\usepackage{makecell} 
\usepackage{tabularx}
\usepackage{chngcntr}
\usepackage[utf8]{inputenc}
\counterwithout{equation}{section}
\counterwithout{figure}{section}
\usepackage[unicode=true,
            pdfusetitle,
            bookmarks=true,
            bookmarksnumbered=false,
            bookmarksopen=false,
            breaklinks=true,
            pdfborder={0 0 0},
            backref=false,
            colorlinks=true,
            hypertexnames=false]{hyperref}
\hypersetup{linkcolor=NavyBlue,urlcolor=NavyBlue,citecolor=NavyBlue}

\begin{document}

\title{Nonlinear dynamics of spatial soliton in a Kerr micro-ring}
\author{Haitao Lv}
\affiliation{College of Science, Hangzhou Dianzi University, Hangzhou, 310018, China}
\author{Chaoying Zhao}
\email{zchy49@163.com}
\affiliation{College of Science, Hangzhou Dianzi University, Hangzhou, 310018, China}
\affiliation{State Key Laboratory of Quantum Optics Technologies and Devices, Shanxi University, Taiyuan, 030006, China}
\date{\today}

\begin{abstract}
The input pump light field can be split into two transverse modes, after entering a AIN micro-ring, which can generate rich nonlinear effects. The cross-phase modulation (XPM) effect in magnetic(TM) polarization mode can cause a refractive index alteration of the micro-ring, the electric(TE) polarization mode and TM polarization mode will display different values and generate a phase change. By adjusting the magnitude of the input TE polarization mode and TM polarization mode, we can achieve a series of phase distributions. By controlling the phase of the electromagnetic field, we can control orbital angular momentum (OAM). The traditional LLE does not take phase into account, in this paper, we obtain a generalized LLE includes phase case. Our research suitable for precision spectroscopy, optical communication links, and coherent information processing. 
\end{abstract}

\maketitle

\section{Introduction}
Chip-scale optical frequency combs generated in high-$Q$ micro-resonators have matured into a versatile platform for broadband coherent light sources with native compatibility to integrated photonics \cite{sun2023microcombs,trocha2018ultrafast}.
By converting a single continuous-wave (CW) pump into a set of mutually coherent lines, micro-combs provide compact multi-wavelength resources for wavelength-division multiplexing, precision metrology, and chip-scale nonlinear and quantum photonics \cite{yu2024tunable,zhu2022modulation,rizzo2023kerrlink,zhang2025nonlinear}.
At the same time, light carrying orbital angular momentum (OAM) offers an independent spatial degree of freedom with an unbounded discrete basis, enabling high-dimensional multiplexing and mode-division processing beyond conventional polarization encoding \cite{allen1992orbital,wang2024integratedSL,zhan2024spatiotemporal}. In experiments, for a comb mode, the frequency domain module performs channel selection and routing. For OAM mode, spatial domain must be engineered and maintained with well-defined routing, mode conversion, and readout \cite{willner2015optical}. A spatial domain module performs OAM conversion on spatially separated fields, with a measurement stage that can validate both intensity and phase for each channel \cite{pirmoradi2025integrated,li2024coam}. 

The input pump light field is split into transverse
electric(TE) and transverse magnetic(TM) modes, the two orthogonally polarized optical fields in the micro-ring can stimulate rich nonlinear effects \cite{zhang2025nonlinear}. The cross-phase modulation (XPM) effect in TM mode can cause a refractive index alteration in micro-ring system, then the TE polarized component and TM polarized component will display a different value. The TE and TM different orthogonally polarized components will cause a change in phase. By continuously adjusting the magnitude of the input TE polarization mode and TM polarization mode, we can achieve continuous phase tuning. By controlling the phase of the electromagnetic field, we can control orbital angular momentum (OAM) of light.  

If we simultaneously possess frequency-domain comb teeth and spatial OAM modes \cite{chen2024integrated,liu2024integrated}, we can make each comb tooth with an OAM order, frequency spatial optical modes suitable for high-capacity communication links, parallel sensing, and multi-channel coherent information processing. 

In this work, a CW pump through waveguide into a high-$Q$ AlN Kerr micro-ring to generate a Kerr micro-comb $\lambda_i$ \cite{pirmoradi2025integrated}. We employ a normalized Kerr-comb model for the micro-ring dynamics \cite{coen2012modeling,zhu2020frequency}. Then we produces a beam with a topological charge $\ell_i$ \cite{marrucci2006spin,biener2002helical,devlin2017arbitrary}. The spatial-mode conversion can be represented as modular operators  \cite{reck1994experimental,clements2016optimal,kaiser2009complete,flamm2012mode}. The complex-field characterization module retrieves tooth-resolved intensity and phase $\{I_i(x,y),\Phi_i(x,y)\}$ \cite{li2024coam,mitsuo1982fourier,leach2004cavity}, the retrieved phase enables direct estimation of the topological charge \cite{willner2015optical,mitsuo1982fourier}. 

\section{THEORETICAL MODEL}

In Fig.1, we adopt AlN micro-ring platform near $1550nm$ with parameters:
ring radius $\simeq 40\mu m$ and TE$_{00}$ waveguide cross-section of
$1.3\mu\ m\times 1.0\mu m$ (AlN thickness $\sim 1.1\mu m$ with a $SiO_2$ over-cladding
$\sim 1.5\mu m$). A photon lifetime on the order of $\kappa^{-1}\sim 0.5ns$. 

\begin{figure}[hptp]
  \centering
\includegraphics[width=1\linewidth]{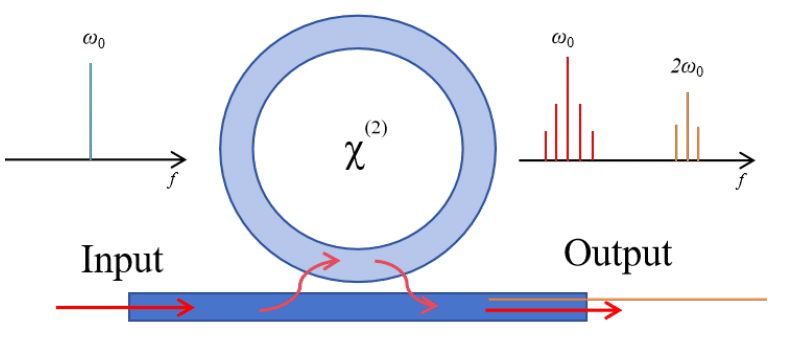}
  \caption{A high-$Q$ AlN micro-ring generation a Kerr micro-comb \cite{pirmoradi2025integrated}}
\end{figure}

The loaded and intrinsic quality factors are taken as $Q_L \approx 6.5\times 10^{5}$ and $Q_{int}\approx 9.3\times 10^{5}$, the cavity linewidth
\begin{equation}
	\kappa/2\pi \approx \frac{c/\lambda_0}{Q_L}\approx 3.0\times 10^{8} Hz
	 (\lambda_0\approx 1550nm),
\end{equation}

The soliton operation is typically matched with on-chip pump powers on the order of $\sim 620mW$.

Without phase, the eigenvalue spectrum of $\hat{\mathcal M}$ determines the MI gain as a function of spatial frequency\cite{zhang2025nonlinear}. With phases $e^{i\ell_\mu\varphi}$, the corresponding eigenvalue spectrum becomes $\hat{\mathcal M}_{\{\ell_\mu\}}$. The conversion-switch parameter $\zeta\in[0,1]$ is implemented through the operator $\hat{\mathcal R}_\zeta$. 
 \begin{equation}
         \hat{\mathcal R}_\zeta \equiv \hat I + \zeta\!\left(\hat{\mathcal P}\hat{\mathcal M}_{\{\ell_\mu\}}-\hat I\right),  
        \end{equation}
 Here, $\mu$ indexes the comb tooth relative to the pump ($\mu=0$).  $\hat{\mathcal P}$ is a propagation operator of micro-ring.
 
The intra-cavity field envelope is described by
\begin{equation}
\begin{split}   
			\frac{\partial E_\zeta(t,\tau;x,y)}{\partial t}
			=
			\hat{\mathcal R}_\zeta\!([
			-1-i\Delta-i\eta\frac{\partial^{2}}{\partial \tau^{2}}\\
			+i\left|\hat{\mathcal R}_\zeta^{-1}E_\zeta(t,\tau;x,y)\right|^{2}]
			\hat{\mathcal R}_\zeta^{-1}E_\zeta(t,\tau;x,y)
			+\sqrt{X}),
         \end{split}
          \end{equation}

 (i)In the case of $\zeta=0$, in the transverse coordinates, 
      \begin{equation} 
	F(t,\tau)=\hat{\mathcal R}_0^{-1}E_0(t,\tau;x,y), 
     \end{equation} 
     
Eq.(3) reduces to the standard intra-cavity LLE \cite{zhu2020frequency,coen2012modeling}. 
\begin{equation}
	\frac{\partial F(t,\tau)}{\partial t}
	=
	\left[
	-1 - i\Delta
	- i\eta\frac{\partial^2}{\partial \tau^2}
	+ i|F(t,\tau)|^2
	\right]F(t,\tau)
	+\sqrt{X}.
\end{equation}
Here $t$ is a dimensionless slow time, $\tau$ is a dimensionless fast time. $\Delta$ is the normalized pump-cavity detuning. $\eta=\pm 1 m^{-1}$ denotes the sign of group-velocity dispersion. For anomalous-dispersion case, we take $\eta=-1 m^{-1}$. $X$ is the normalized pump power. 

For a CW solution
\begin{equation}
F(t,\tau)=F_s,
\end{equation}
the intra-cavity intensity satisfies the bistability relation
\begin{equation}
Y=|F_s|^2, X = Y^3 - 2\Delta Y^2 + (\Delta^2+1)Y.
\end{equation}
 A weak complex noise seed is added to the initial CW background to trigger MI sideband growth.
\begin{equation}
    F(t,\tau)=F_s + a(t,\tau),
\end{equation}
After normalization, $u$ and $v$ are the
steady-state solutions of the two optical pulses
\begin{equation}
    a(t,\tau)=u\,e^{\lambda t+i\Omega\tau}+ve^{\lambda t-i\Omega\tau},
\end{equation}
Here, $u$ and $v$ are small eigenmode coefficients of the perturbation sidebands at $+\Omega$ and $-\Omega$, respectively. $\Omega$ is the perturbation frequency in the fast-time domain. 
Linearizing Eq.(9) yields the eigenvalues
\begin{equation}
	\lambda_{\pm}(\Omega)
	=
	-1
	\pm
	\sqrt{
		Y^2-\left(2Y-\Delta+\eta\Omega^2\right)^2
	},
\end{equation}
The MI gain is defined as
\begin{equation}
g(\Omega)=\max\{0,\mathrm{Re}[\lambda_{+}(\Omega)]\}. 
\end{equation}
where positive-gain bands provide the initial sidebands that seed comb formation.

Fig.2 summarizes the MI analysis and the CW branches at the operating parameters. In Fig.2(a), the shaded bands with $g>0$ indicate the unstable mode ranges that initiate sideband growth. Fig.2(b) shows the multi-valued CW response $Y(\Delta)$ at fixed pump power $X$, which guides the detuning-scan trajectory employed in the time-domain split-step simulations.
\begin{figure*}[htbp]
\centering
\includegraphics[width=1\linewidth]{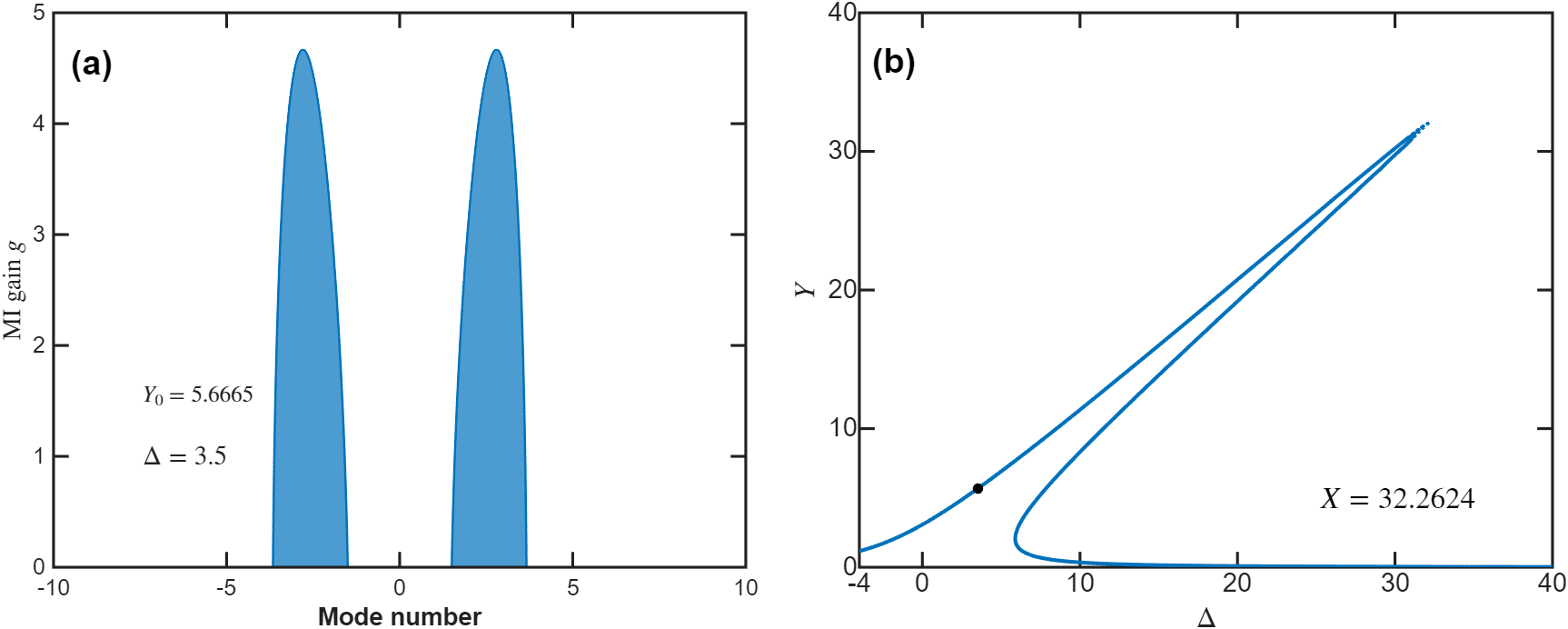}
\caption{(a) AlN micro-ring near $1550nm$ with parameters are as follows: radius: $\simeq 40\mu m$, thickness: $\sim 1.1\mu m$, $SiO_2$ over-cladding: $\sim 1.5\mu m$, cross section area: $1.3\mu\ m \times 1.0\mu m$. Modulation-instability (MI) gain spectrum $g$ versus mode number $\mu$
at the operating point $\Delta=3.5 m^{-1}$ with the CW background intensity $Y_0=5.6665 mW^2$ (anomalous-dispersion). The shaded regions indicate mode bands with positive gain that seed initial comb sidebands. (b) Continuous-wave (CW) bistability curve $Y$ as a function of detuning $\Delta$ at fixed normalized pump power $X=32.2624 mW^2$, showing the multi-valued steady-state branches used to identify the detuning-scan path toward soliton formation.}
\end{figure*}

\begin{figure*}[htbp]
	\centering
\includegraphics[width=1\linewidth]{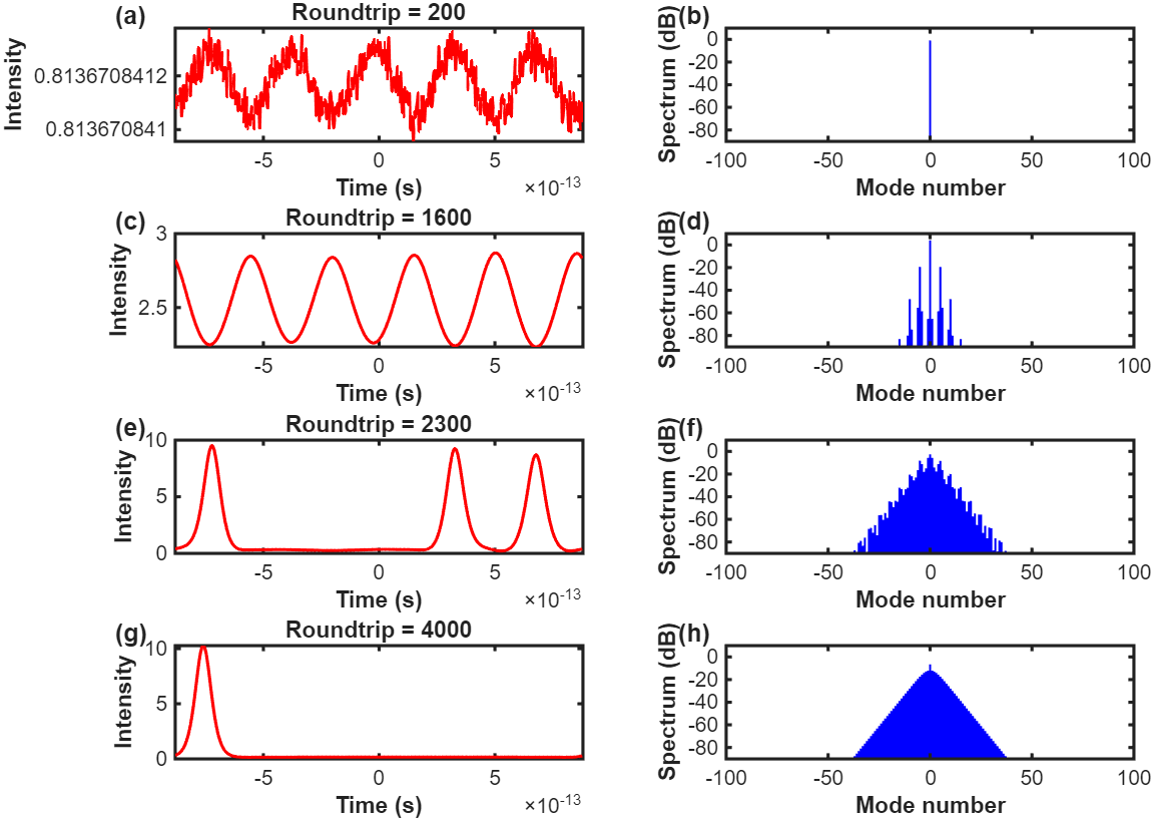}
    \caption{Evolution of the intra-cavity field at slow-time steps (round-trips).
	Left column: temporal intensity profiles $|F(\tau)|^{2}$ in the fast-time domain; right column: corresponding optical spectra in dB versus mode number.
	(a,b) Roundtrip $=200$: CW background with small fluctuations and a nearly single-line spectrum.
	(c,d) Roundtrip $=1600$: emergence of a periodic modulation (MI/Turing pattern) accompanied by initial sidebands.
	(e,f) Roundtrip $=2300$: multi-soliton state with multiple pulses in the time domain and a broadened comb spectrum.
	(g,h) Roundtrip $=4000$: a stable single-soliton state featuring a clean isolated pulse and a smooth, broadband comb envelope.}
\end{figure*}

\begin{figure*}[htbp]
	\centering
\includegraphics[width=1\linewidth]{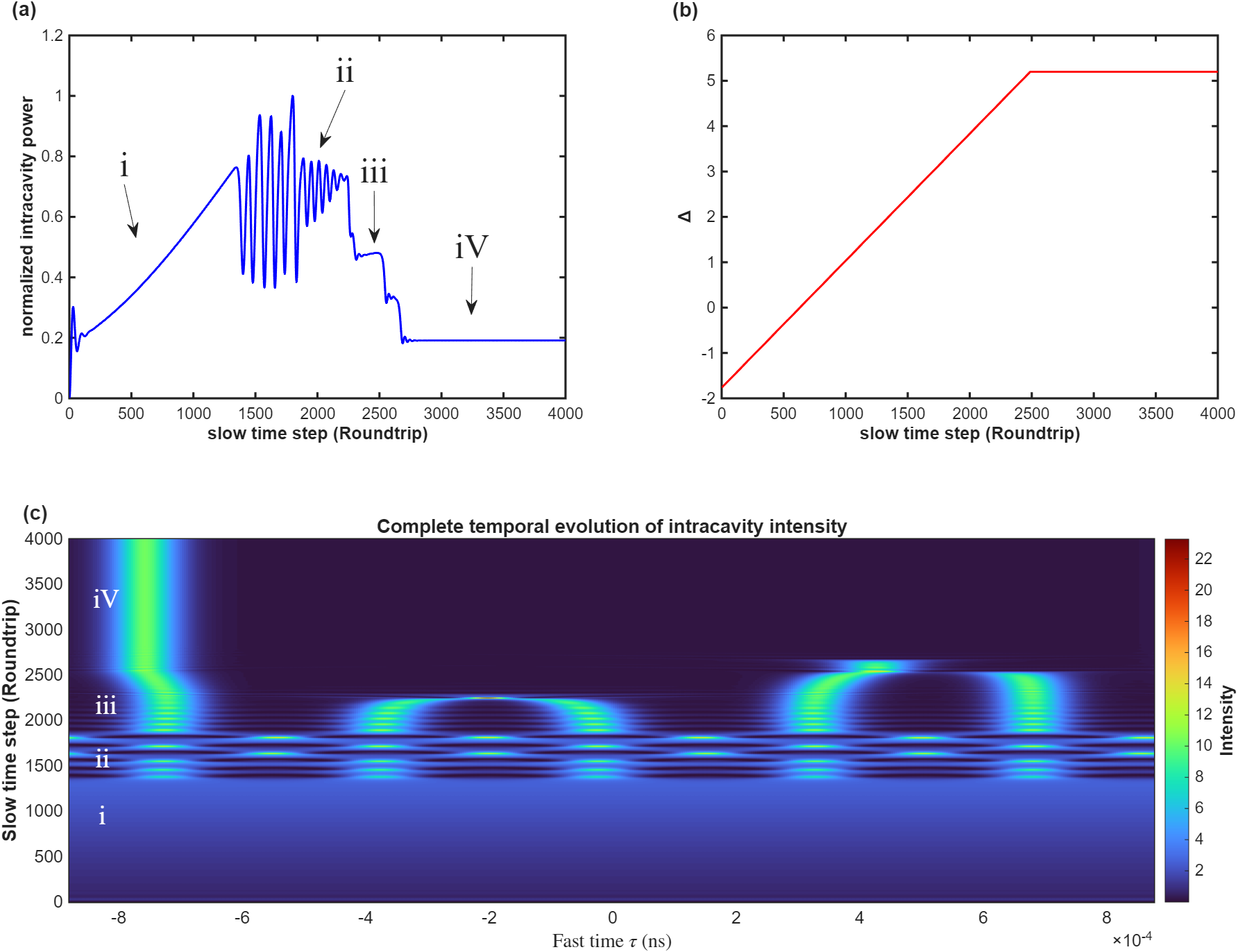}
    \caption{Temporal evolution of Kerr-comb dynamics during a pump-cavity detuning scan.
	(a) Normalized intra-cavity power versus slow-time step (roundtrip), showing the transition from a CW background to an unstable regime and subsequent step-like soliton formation.
	(b) The programmed detuning $\Delta$ as a function of slow time (linear sweep followed by a hold).
	(c) Spatiotemporal map of the intra-cavity intensity $|F(\tau,t)|^{2}$ over the fast time $\tau$ and the slow time $t$.
	The dynamical stages including: (i) CW buildup, (ii) modulational-instability-driven patterns/chaos, (iii) soliton switching with stepwise changes in the average intra-cavity power, and (iv) a final stationary soliton state.}
\end{figure*}

Fig.3 presents snapshots of the intra-cavity field at selected round-trips during the scan.
At step $=200$, the intra-cavity waveform remains close to the CW background with small fluctuations, and the spectrum is dominated by the pump line. At roundtrip=1600, MI side-bands development and a Turing pattern emerge, accompanied by multiple discrete comb lines around the pump. As the detuning is further increased, localized pulses appear at roundtrip=2300, indicating the onset of soliton formation with multi-soliton or soliton-crystal-like states.
Finally, at roundtrip=4000, the system converges to a stable single-soliton state, evidenced by an isolated pulse in the fast-time domain and a smooth, broadband comb envelope in the spectral domain.

We now turn to time-domain simulations of Eq.(2) to capture the full nonlinear evolution from the CW background to MI, ultimately to dissipative Kerr solitons. The normalized LLE is integrated using a split-step Fourier method on a discrete mode grid of $N=2^{9}$ points in the fast-time window, corresponding to $N$ resonator modes indexed by the relative mode number $\mu$.
To deterministically reach the soliton-supporting regime, we apply a detuning scan $\Delta$, where $q$ denotes the slow-time step. $\Delta$ is ramped from a blue-detuned value to a red-detuned value and then held constant,
\begin{equation}
	\Delta=
	\begin{cases}
\Delta_{\min}+\frac{\Delta_{\max}-\Delta_{\min}}{q_{r}}q, & 0\le q\le q_{r},\\
		\Delta_{\max}, & q>q_{r},
	\end{cases}
\end{equation}
with $\Delta_{\min}\approx -1.8 m^{-1}$, $\Delta_{\max}\approx 5.2 m^{-1}$, and a linear detuning ramp up to roundtrip $\approx 2500$, as shown in Fig.4(b).
The intra-cavity response is characterized by the normalized intra-cavity power
$\langle |F(t,\tau)|^{2}\rangle_{\tau}$ plotted in Fig.4(a), while the complete spatiotemporal evolution $|F(t,\tau)|^{2}$ is summarized in Fig.4(c).

 In stage ii, corresponding to the MI/chaotic region, the intra-cavity power exhibits strong oscillation and the spatiotemporal map shows extended pattern dynamics. In stage iii, step-like drops occur when the system switches between distinct attractors, notably during soliton-number reduction. After the scanning reaches and holds $\Delta_{\max}$, the intra-cavity power stabilizes, consistent with a stationary single-soliton solution.

(ii) In the case of $\zeta=1$, we have
\begin{equation}
     \hat{\mathcal R}_1^{-1} E_{1}(t,\tau;x,y)= \hat{\mathcal P}\hat{\mathcal M}_{\{\ell_\mu\}}F(t,\tau). 
 \end{equation}     
The output field after OAM tagging. 
	\begin{equation}
		\hat{\mathcal M}_{\{\ell_\mu\}}
		=
    \sum_{\mu}\hat{\Pi}_{\mu}\otimes \hat U_{\ell_\mu},
	\end{equation}
where $\hat{\Pi}_{\mu}$ projects onto the $\mu th$ comb tooth in frequency space and $\hat U_{\ell_\mu}$ acts on the transverse field. 

The intra-cavity field is expanded in angular modes
\begin{equation}
A(\vartheta,t)=\sum_{\mu} a_\mu(t)e^{i\mu\vartheta},
\end{equation}
where $\vartheta$ denotes the micro-ring resonator angular coordinate. 
Using the standard input-output relation for a side-coupled ring, 
\begin{equation}
	s_{out,\mu}(t)=\sqrt{\kappa_{ex}}a_\mu(t)-s_{in}\delta_{\mu,0},
\end{equation}
which provides the set of comb-tooth complex amplitudes $s_{out,\mu}$. 

The soliton micro-comb provides a set of discrete frequencies $\omega_\mu$. 
The corresponding tooth frequency is
\begin{equation}
	\omega_\mu=\omega_0 + D_1\mu + \tfrac{1}{2}D_2\mu^2+\cdots.
\end{equation}
where $D_1$ is the (angular) free-spectral range and $D_2$ is the second-order dispersion coefficient.
For the AlN micro-ring, we adopt $D_1/2\pi \approx 0.6THz$ and $D_2/2\pi \approx 40MHz$.   
\begin{equation}
(\hat U_{\ell_\mu}E)(x,y)=A_k(x,y)
	E(x,y)e^{i\phi(x,y;\sigma)},
\end{equation}
 The field of tooth $\omega_\mu$ is modeled as
\begin{equation}
E_{\mu}(x,y)=\mathcal{A}_\mu u(x,y)
		e^{i\phi(x,y;\sigma)}.
\end{equation}
where $\mathcal{A}_\mu\propto s_{\text{out},\mu}$ carries the tooth amplitude and phase, $u(x,y)$ is the
 envelope of $A_k(x,y)$.

The propagation operator after OAM tagging.
	\begin{equation}
		\hat{\mathcal P}
		=
	\sum_{\mu}\hat{\Pi}_{\mu}\otimes \hat{\mathcal P}_{\lambda_\mu}(z),
	\end{equation}
where $\hat{\mathcal P}_{\lambda_\mu}(z)$ is evaluated at each tooth wavelength $\lambda_\mu$.
We implement $\hat{\mathcal P}_{\lambda_\mu}(z)$ by using the angular-spectrum method
\begin{equation}
		(\hat{\mathcal P}_{\lambda_\mu}(z)E)(x,y)
		=
		\mathcal{F}^{-1}\!\left\{
		\mathcal{F}\{E(x,y)\}
		H_\mu(k_x,k_y;z)
		\right\},
	\end{equation}
with the transfer function
	\begin{equation}
		H_\mu(k_x,k_y;z)
		=
		e^{i z\sqrt{k_\mu^{2}-k_x^{2}-k_y^{2}}}
		\Theta\!\left(k_\mu^{2}-k_x^{2}-k_y^{2}\right),
	\end{equation}
The wavelength-dependent scaling
    \begin{equation}
				k_\mu=\frac{2\pi}{\lambda_\mu}.
	\end{equation}

If the detection averages out inter-tooth beating, the measured intensity is well approximated by an incoherent sum. The tooth-resolved intensity and phase are
\begin{equation}
I_{\mathrm{tot}}\approx \sum_{\mu} I_\mu=\sum_{\mu}|\mathcal{F}\{E_\mu(x,y)\}|^2,
\end{equation}
\begin{equation}
\Phi_\mu=\arg[\mathcal{F}\{E_\mu(x,y)\}].
\end{equation}

\section{NUMERICAL SIMULATION}

As an overview, Fig.5 summarizes the chaotic teeth-OAM conversion for a chaotic-state comb snapshot.

\begin{figure*}[htbp]
  \centering
\includegraphics[width=1\textwidth]{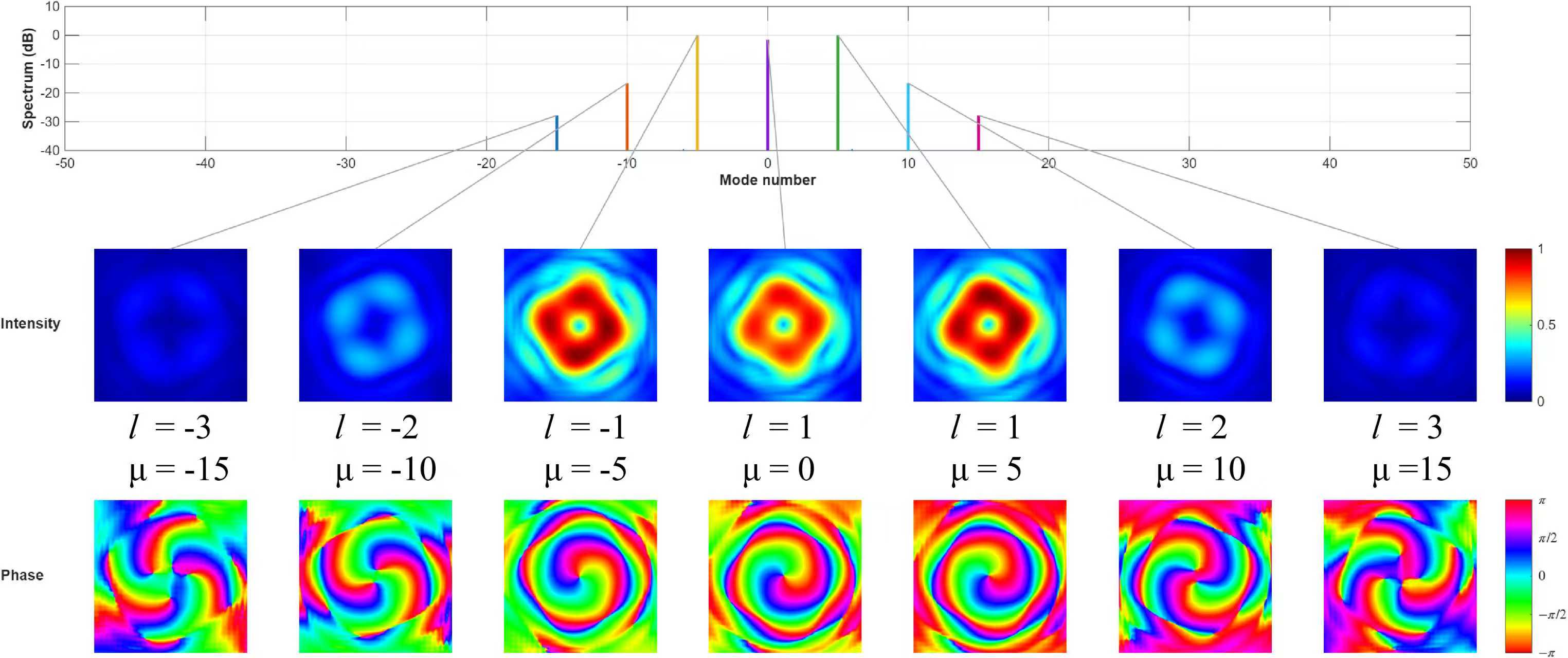}
  \caption{ Overview of chaotic teeth-OAM conversion. Top: selected comb lines (a chaotic-state snapshot). Middle: intensity distributions. Bottom: phase profiles with topological charges $\ell=\{-3,-2,-1,1,1,2,3\}$ and comb-tooth modes $\mu=\{-15,-10,-5,0,5,10,15\}$.}
\end{figure*}

Fig.5 provides a view of the $\zeta=1$ readout: comb teeth (top) converted into intensity and phase are shown in the middle and bottom rows, respectively. In the chaotic-state snapshot, we choose seven charges $\ell=\{-3,-2,-1,1,1,2,3\}$ and the comb-tooth mode $\mu=\{-15,-10,-5,0,5,10,15\}$, which demonstrate that the phase on each wavelength when the comb amplitudes are irregular. The intensity maps exhibit annular profiles with a suppressed on-axis field, while the wrapped phase maps show $|\ell|$ azimuthal $2\pi$ windings around the beam center. 
    
    \begin{figure*}[htbp]
  \centering
\includegraphics[width=1\textwidth]{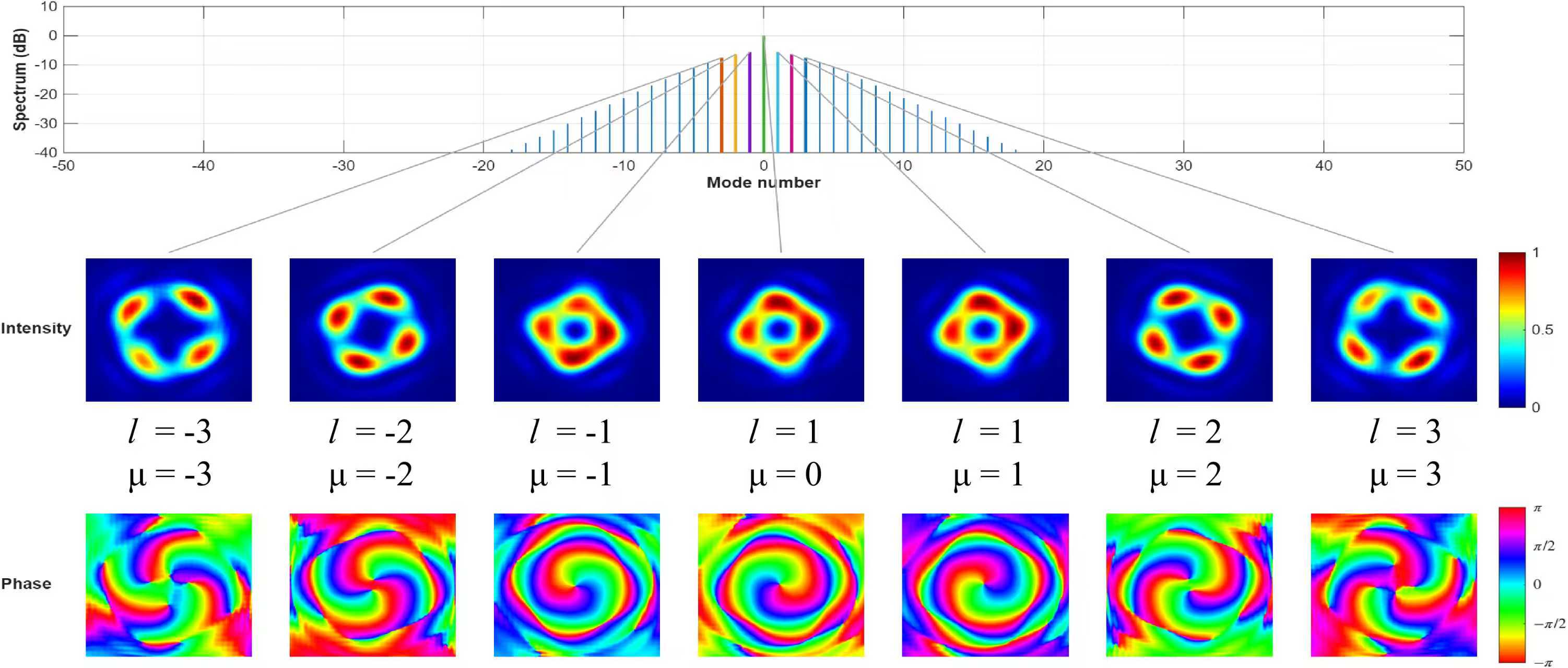}
  \caption{Tooth-OAM micro-comb ($\zeta=1$). Top: simulated micro-comb spectrum with the selected comb teeth highlighted. Bottom: focal-plane intensity and wrapped phase, demonstrating the assigned OAM orders $\ell=\{-3,-2,-1,1,1,2,3\}$ and the comb-tooth mode $\mu=\{-3,-2,-1,0,1,2,3\}$.}
\end{figure*}

In Fig.6, the top panel shows the simulated micro-comb spectrum together with the subset of comb teeth and OAM orders. We program $\ell=\{-3,-2,-1,1,1,2,3\}$ and the comb-tooth mode $\mu=\{-3,-2,-1,0,1,2,3\}$ for seven teeth, illustrating that distinct wavelength can be mapped to either distinct or identical OAM charges. The lower panels display the intensity $I_\mu$ and wrapped phase $\Phi_\mu$, which exhibit annular profiles and azimuthal phase windings. 
    
Each tooth can be represented by a Jones matrix \cite{wang2023metasurface} in the linear basis $\{H,V\}$. 
\begin{equation}
\mathbf{E}_{\mu}^{k}=
\begin{pmatrix}
E_{\mu,H}^{k}\\
E_{\mu,V}^{k}
\end{pmatrix}=\sum_{\mu'} T_{k\mu'}(\omega_{\mu'})\mathbf{E}_{\mu'},
\end{equation}
Here, the summation over $\mu'$ accounts for the  nonzero leakage/crosstalk between neighboring teeth. In the ideal limit, the transfer matrix $T_{k\mu}\approx \delta_{k\mu}$.

We switch basis from $\{H,V\}$ to basis $\{\sigma=\pm1\}$ 
\begin{equation}
\begin{pmatrix}
E_{\mu,+}^{k}\\
E_{\mu,-}^{k}
\end{pmatrix}
=
\frac{1}{\sqrt{2}}
\begin{pmatrix}
1 & i\\
1 & -i
\end{pmatrix}
\begin{pmatrix}
E_{\mu,H}^{k}\\
E_{\mu,V}^{k}
\end{pmatrix}.
\end{equation}
The phase is
 \begin{equation}
\phi(x,y;\sigma)=2\sigma\theta_k(x,y), 
 \end{equation}
where $\theta_k(x,y)$ is the local in-plane orientation. In order to evaluate OAM, we switch from Cartesian coordinates $(x,y)$ to polar coordinates $(\rho,\varphi)$ via
 \begin{equation}
x=\rho\cos[\varphi], y=\rho\sin[\varphi].
 \end{equation}
We then sampling the complex field on a circle of radius $\rho=\rho_0$ (the radius of maximum micro-ring intensity) and perform an angular along $\varphi\in[0,2\pi]$. 
\begin{equation}
\theta(\rho,\varphi)=\frac{\ell_\mu}{2}\,\varphi+\theta_{0},
\end{equation}
where $\theta_0$ contributes a spatial phase factor $e^{i2\sigma\theta_0}$ and therefore does not change the OAM charge. The generated beam with topological charge $\ell_\mu$ for each tooth, the standard azimuthal orientation profile yields an azimuthal phase factor $e^{i\ell\varphi}$ with $\ell=\sigma\ell_\mu$. 

We perform a decomposition on a ring of radius $\rho_0$. 
\begin{equation}
	c_{\mu,\ell}(\rho_0)=\frac{1}{2\pi}\int_{0}^{2\pi} 
	E_\mu(\rho_0,\varphi)e^{-i\ell\varphi}d\varphi,
\end{equation}
and the normalized OAM power distribution
\begin{equation}
\mathcal{S}_\mu(\ell)=\frac{|c_{\mu,\ell}(\rho_0)|^2}{\sum_{\ell'}|c_{\mu,\ell'}(\rho_0)|^2}.
\end{equation}

\begin{figure*}[htbp]
  \centering
\includegraphics[width=1\linewidth]{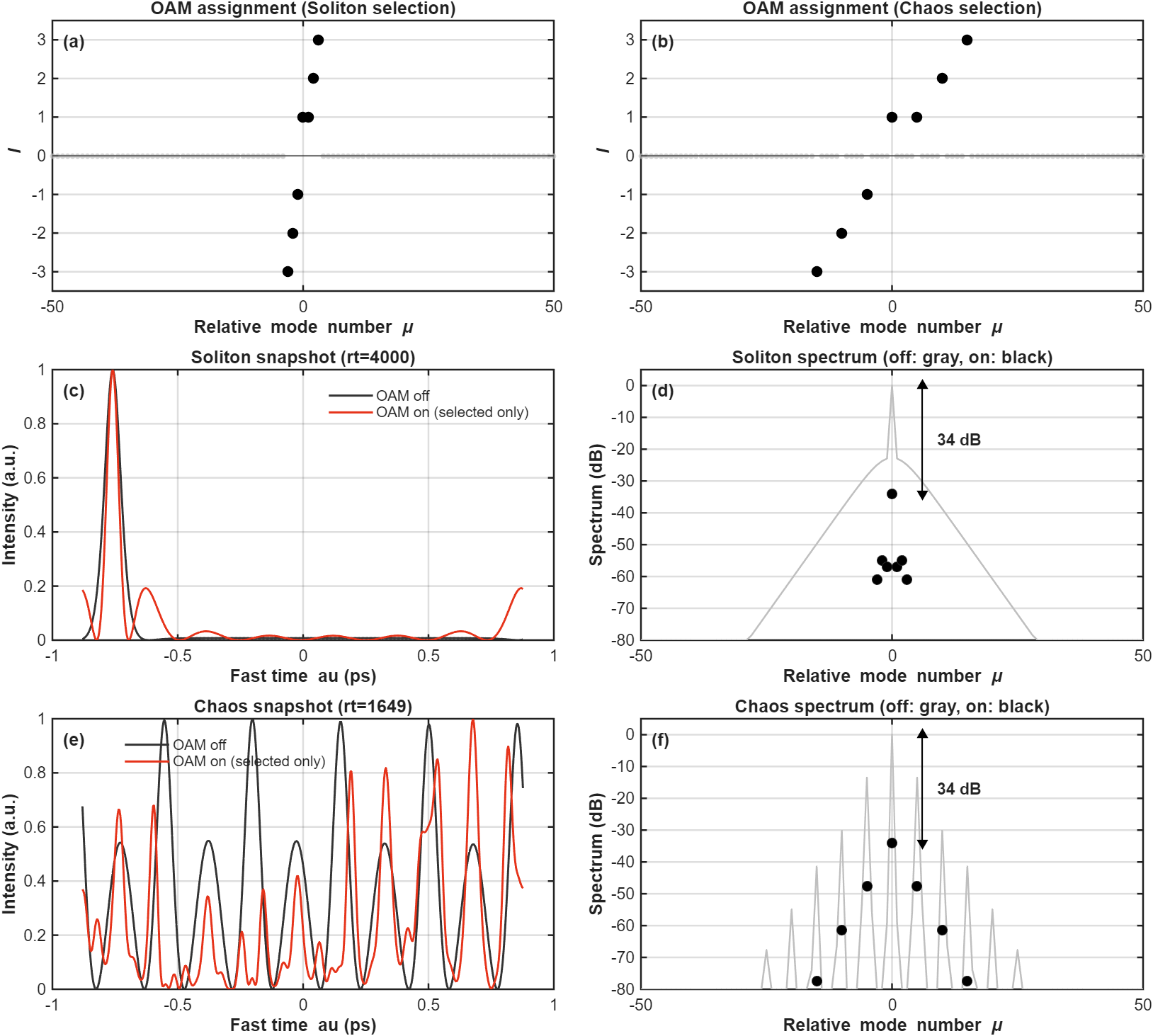}
\caption{Teeth-OAM for soliton and chaotic comb states. (a,b)Tooth-to-$\ell$ for the soliton and chaos selections (unselected teeth are shown on the $\ell=0$ baseline). (c,e) Time-domain snapshots reconstructed from the full comb (without OAM, black line) and from the tooth subset used for OAM tagging (with OAM, red line). (d,f) Corresponding spectra: the gray curve shows the full comb spectrum (without OAM), while black circles mark the teeth-OAM reconstruction. The arrow indicates the power contrast $34dB$ between the pump line and the tooth level.}
\end{figure*}

\section{Conclusion}
Our numerical simulations demonstrate how the OAM-tagged multi-tooth field evolves after conversion and provide an intuitive link between OAM mapping and temporal readout. We consider without OAM case and with OAM case for soliton and chaotic comb states, respectively. In the case of OAM, only a small subset of comb teeth are non-zero charges (with $\mu=0$ enforced as $\ell=1$ and the remaining selected teeth following the mapping), while all unselected teeth are set to $\ell=0$. 

\begin{acknowledgments}
This work was funded by the State Key Laboratory of Quantum Optics Technologies and Devices, Shanxi University, Shanxi, China (Grants No.KF202503).
\end{acknowledgments}

\section*{DATA AVAILABILITY}
The data that support the findings of this article are not publicly available. The data are available from the authors upon reasonable request.

\bibliography{ref}

\end{document}